\documentclass[reprint,
superscriptaddress,
showpacs,
 amsmath,amssymb,
 aps,
]{revtex4-1}

\usepackage{graphicx}
\usepackage{hyperref}
\usepackage{color}

\renewcommand\Re{{\rm Re}}

\newcommand{\ket}[1]{|{#1}\rangle}

\begin{document}

\preprint{APS/Non-harmonic}

\title{Suppression of Rabi oscillations in hybrid optomechanical systems}

\author{Timo Holz}
\affiliation{Theoretische Physik, Universit\"at des Saarlandes, D-66123 Saarbr\"ucken, Germany}
\author{Ralf Betzholz}
\email{ralf.betzholz@physik.uni-saarland.de}
\affiliation{Theoretische Physik, Universit\"at des Saarlandes, D-66123 Saarbr\"ucken, Germany}
\author{Marc Bienert}
\affiliation{Theoretische Physik, Universit\"at des Saarlandes, D-66123 Saarbr\"ucken, Germany}
\affiliation{Hohenzollern Gymnasium, D-72488 Sigmaringen, Germany}

\date{\today}


\begin{abstract}
In a hybrid optomechanical setup consisting of a two-level atom in a cavity with a pendular end-mirror, the interplay between the light field's radiation pressure on the mirror and the dipole interaction with the atom can lead to an effect, which manifests itself in the suppression of Rabi oscillations of the atomic population. This effect is present when the system is in the single-photon strong coupling regime and has an analogy in the photon blockade of optomechanics.
\end{abstract}

\maketitle

\section{Introduction}
In the Jaynes-Cummings model, a single atomic dipole interacting with a single photon of an electromagnetic field mode in a cavity periodically exchanges the excitation between the electronic and photonic degree of freedom. In this letter we ask ourselves: How does this behavior change, when the cavity's mirrors are not fixed, but can move in time as an additional quantum dynamical degree of freedom? How does the quantumness of the mirror motion manifest itself in the dynamics? 
And, what happens if several photons are present in the cavity field? 

The recent experimental success in cavity optomechanics~\cite{schliesser2008,chan2011,camerer2011,aspelmeyer2014}, notably the demonstration of laser cooling of the cavity's mirror motion towards the ground state of the confining harmonic potential, pioneers near-future experiments with extended, hybrid systems. An obvious extension of the paradigmatic optomechanical system, i.e an optical cavity coupled to a mechanical element by radiation forces whose motion changes the cavity's boundaries, can be achieved by adding a single two-level system which couples to the cavity light field by dipole interaction, thereby combining cavity optomechanics with cavity quantum electrodynamics. Besides conventional quantum electrodynamics experiments with macroscopic cavities investigations of such systems are of particular importance for solid-state systems, such as two-level systems in optomechanical crystals~\cite{eichenfield2009,riedrich-moller2012,breyer2012}. In circuit electromechanical setups the coupling to two-level systems has already been demonstrated~\cite{lahaye2009,pirkkalainen2013,lecocq2015,pirkkalainen2015}.

Especially solid-state systems (leaving aside the two-level system in contemporary realizations) are candidates for reaching the so-called single-photon strong coupling regime~\cite{nunnenkamp2011}, where the radiation force of a single photon is strong enough to displace the harmonically oscillating mechanical element by more than the extension of its ground state wave packet. This regime is entered, when the mechanical frequency becomes much smaller than the optomechanical coupling. Then, the non-linearity of the radiation force becomes clearly apparent and manifests itself in phenomena like the photon blockade effect~\cite{rabl2011}, which is caused by an effective photon-photon interaction mediated by the mechanical degree of freedom.

When adding a single atom to the optomechanical cavity, the resulting hybrid system does not only reflect the radiation forces of the cavity-mirror interaction, but also includes the dipole interaction between a near-resonant atomic transition and the cavity field, drastically altering the dynamics as a whole. The richness in the dynamics of such a highly non-linear system has been started to be explored in several general investigations of its coherent motion~\cite{wang2008,chang2009,ramos2013} and can be exploited to prepare non-classical states, similar as in nonlinear optical setups~\cite{poyatos1996,ritsch2004}. In this letter we focus on the possibility to dynamically suppress the Rabi oscillation of the atomic population due to the mirror motion. The idea behind can best be illustrated having a classical picture in mind, i.e. when the mirror is moving parametrically along a given trajectory. Then, an initially tuned atom-cavity interaction becomes disturbed by the mirror motion due to the change of the cavity's resonance frequency being a function of the cavity length. If the mirror elongation is large enough, the modified atom-cavity detuning can lead to reduced Rabi oscillations of the quantum electrodynamic subsystem. We will develop this simple classical picture beyond the trivial parametric motion towards a fully dynamical one in the quantum regime. The Wigner function will be used to represent the dynamics of the mechanical degree of freedom, which can exhibit strong non-classical characteristics deep in the non-linear parameter regime. Even when dissipation and variation of the initial state is included, characteristic features of the suppression can be distinguished.

\section{The hybrid optomechanical setup}
\begin{figure}[h!]
\begin{center}
\includegraphics[width=0.6\linewidth]{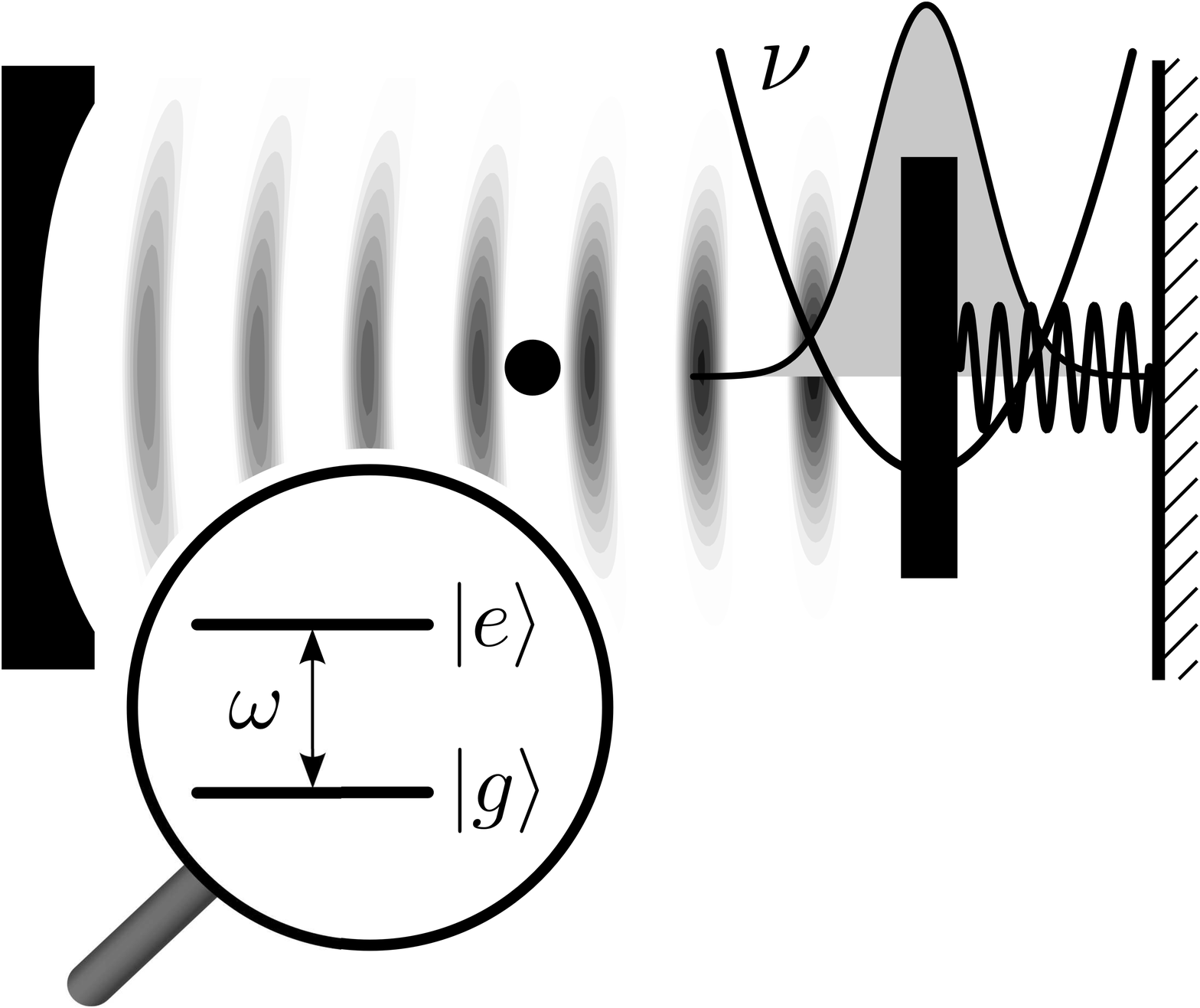}
\end{center}
\caption{\label{fig.system}Hybrid optomechanical setup consisting of a single-mode cavity with a moving end-mirror confined in a harmonic potential of frequency $\nu$ and a resonant two-level atom at a fixed position. The cavity couples to the mechanical oscillator by radiation pressure with a coupling strength $\chi$ and to the two-level atom via dipole interaction of strength $g$.	}
\end{figure}

We consider an optomechanical resonator with a two-level atom placed at a fixed position within the resonator. The transition frequency between the ground state $\vert g\rangle$ and the excited state $\vert e\rangle$ is assumed to be resonant with the frequency $\omega$ of the cavity when its end mirror is in its equilibrium position. The atomic dipole couples via 
Jaynes-Cummings interaction~\cite{jaynes1963} to the electromagnetic field of the cavity while the cavity interacts with the mechanical oscillator of frequency $\nu$ and mass $M$ by radiation pressure. We denote the dipole-cavity interaction strength and the optomechanical coupling strength by $g$ and $\chi$, respectively. A schematic picture of this hybrid quantum system is shown in Fig.~\ref{fig.system}.
The total Hamiltonian of the hybrid systems reads
\begin{align}
\label{eq.hamiltonian1}H = H_{\rm om}+H_{\rm tls}+H_{\rm JC}
\end{align}
where the operator $H_{\rm om}$ labels the standard Hamiltonian of optomechanics~\cite{law1995}
\begin{align}
\label{eq.optohamiltonian}
H_{\rm om}=\hbar\omega a^\dagger a+\hbar\nu b^\dagger b -\hbar\chi a^\dagger a \left( b+b^\dagger\right)\,.
\end{align}
Here, $a$ and $a^\dagger$ represent annihilation and creation operators of the cavity mode, and $b$ and $b^\dagger$ for the mechanical oscillator. Its position operator relative to the equilibrium is $x=\xi(b+b^\dagger)$ where the harmonic oscillator length scale $\xi=\sqrt{\hbar/2M\nu}$ denotes the extension of the ground state wave packet. The Hamiltonians of the two-level system and the Jaynes-Cummings interaction are given by
\begin{align}
\label{eq.hamiltonian2tls}
H_{\rm tls} &= \hbar\omega\vert e\rangle\!\langle e\vert \, , \\
\label{eq.hamiltonian2JC}
H_{\rm JC} &= \hbar g\left[a^\dagger\vert g\rangle\!\langle e\vert+a\vert e\rangle\!\langle g\vert\right]\,,
\end{align}
respectively.

\section{Dynamics of the optomechanical subsystem}
\begin{figure}[h!]
\flushleft{\hspace{8ex} a)}\vspace{-4ex}
\begin{center}
\includegraphics[width=0.65\linewidth]{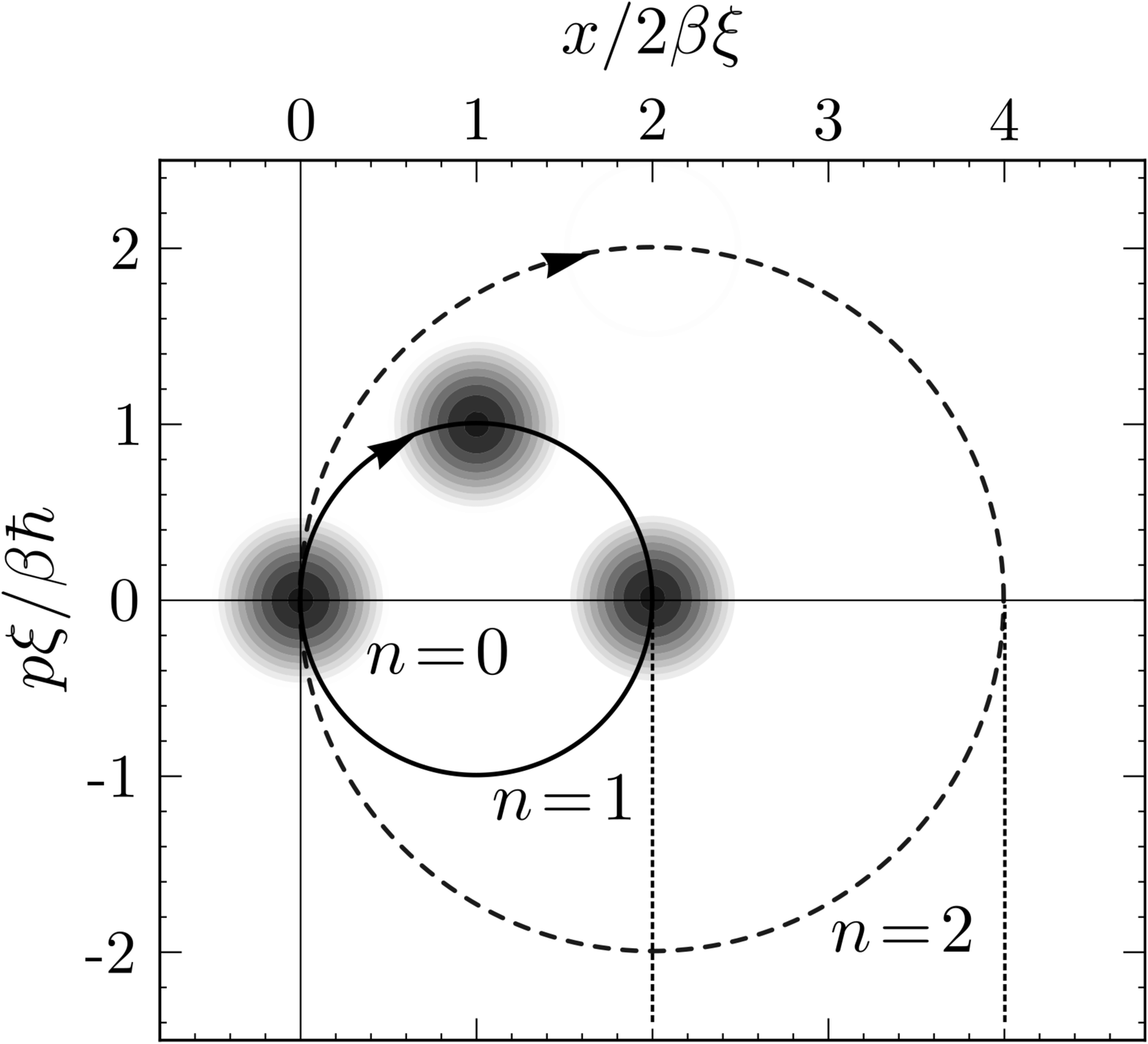}
\end{center}\vspace{-4.5ex}
\flushleft{\hspace{8ex} b)}\vspace{-6ex}
\begin{center}
\includegraphics[width=0.65\linewidth]{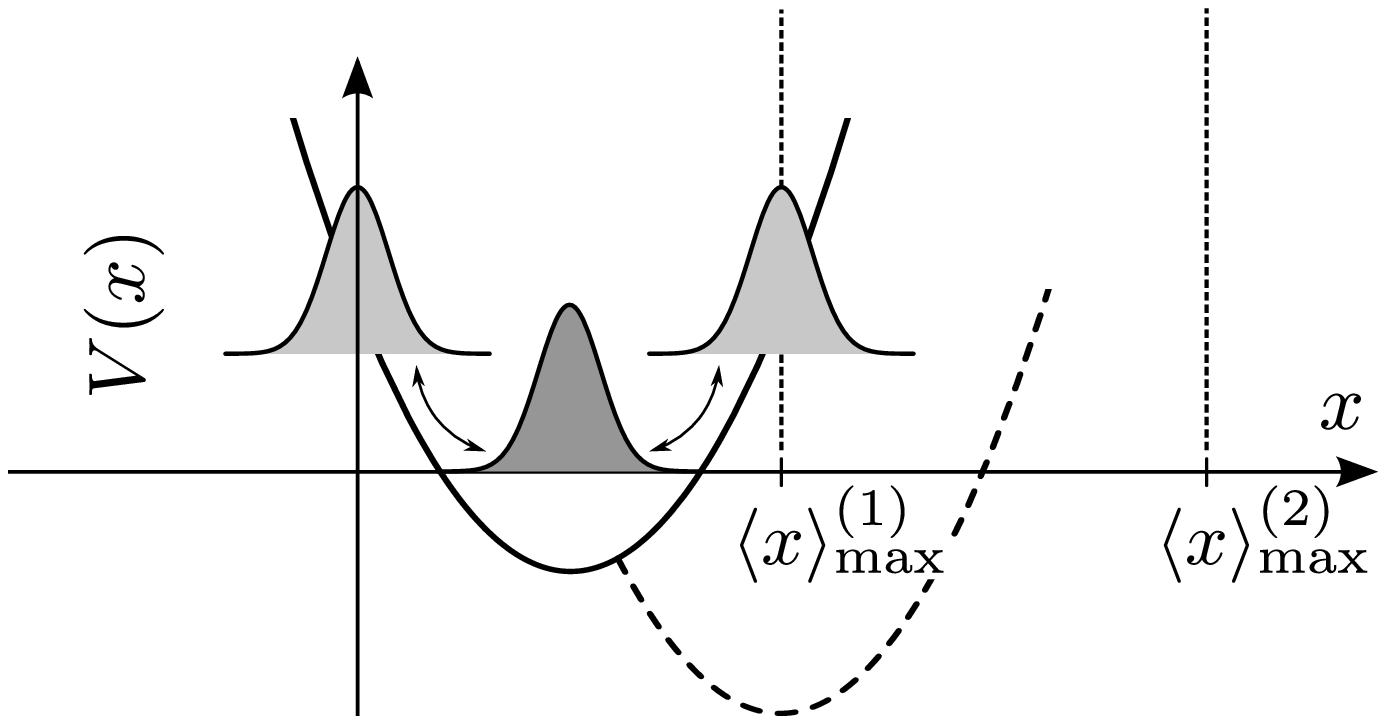}
\end{center}
\caption{\label{fig.trajectories}a) Circular trajectories of the mechanical oscillator's Wigner function for cavity photon numbers $n=1$ (solid) and $n=2$ (dashed) if the mirror was initially prepared in its ground state. The radii are proportional to $\beta$ while for $n=0$ the Wigner function rests centered around the origin. b) Trajectory of the oscillator wave packet in the harmonic potential modified by the radiation pressure. The $\langle x\rangle^{(n)}_{\rm max}$ mark the maximal elongations.}
\end{figure}

Before we delve into the motion of the full hybrid system we set the stage by recalling the dynamics of the optomechanical system in absence of the atom. The eigenstates of $H_{\rm om}$ are given by~\cite{bose1997}
\begin{equation}
\vert n \rangle\, D(n\beta)\vert m\rangle_{\rm mec}
\end{equation}
with the definition $\beta=\chi/\nu$ and displacement operators $D(\alpha)=\exp[\alpha b^\dagger-\alpha^\ast b]$. The 
states $\vert n \rangle$ and $\vert m\rangle_{\rm mec}$ stand for the cavity and mechanical Fock states, respectively. 
The propagation of the mechanical oscillator, initially in prepared in its ground state, is given by the coherent state~\cite{bose1997}
\begin{align}
\vert\eta(t)\rangle_{\rm mec}=\vert n\beta(1-e^{-i\nu t})\rangle_{\rm mec}\,.
\end{align} 
Figure~\ref{fig.trajectories}a) shows the trajectories in Wigner phase space for the values $n=1$ and $n=2$ of the cavity photon number. 
The radiation pressure pushes the mirror out of its equilibrium position, resulting in a modified cavity length and thereby a modified frequency. The maximum displacement from the equilibrium position
\begin{align}
\label{eq.displacement}
\langle x\rangle_{\rm max}^{(n)}=2\xi\,\Re[n\beta(1-e^{-i\nu t})]= 4n\beta\xi
\end{align}
is reached at time $t=\pi/\nu$. In the single-photon strong coupling regime this displacement amounts to a significant change in the cavity frequency during the dynamics.

\section{Coherent dynamics of the hybrid system}
For the coherent dynamics of the hybrid system including cavity, atom and mechanical oscillator, we focus on the initial state $\vert\psi(t=0)\rangle=\vert e\rangle\vert 0\rangle\vert 0\rangle_{\rm mec}$. Experimentally such a state can be prepared with the help of ground-state cooling of the mechanical oscillator and standard optical pumping for the atomic state. Ground-state cooling for an optomechanical crystal was already demonstrated~\cite{chan2011} and is expected to be realizable in similar hybrid systems in near-future experiments.  

The position of the mirror determines the frequency of the cavity, which comes out most clearly when rewriting the Hamiltonian~\eqref{eq.hamiltonian1} in the form
\begin{align}
\label{eq.hamiltonian2}
H=\hbar\left[\omega-\frac{\chi}{\xi}x\right]a^\dagger a+\hbar\nu b^\dagger b+H_{\rm tls}+H_{\rm JC}\,.
\end{align}
The mechanical oscillator in its maximally elongated mean position Eq.~\eqref{eq.displacement} results in an effective detuning $\bar\delta_n=4n\beta\chi$ between the cavity and the atom. From this consideration follows that for a distinct effect of the mirror's motion on the Rabi oscillation one has to be in the strong-coupling regime with $\beta\gtrsim 1$. In this regime we cannot resort to the approximate solution of single polariton optomechanics~\cite{restrepo2014}. 

\subsection{Limiting cases}
Nevertheless, to provide first insight into the dynamics we start our investigation with two limiting cases, namely $g\ll\chi,\nu$ and $\beta\ll 1$, that allow for an approximate analytical treatment. 
\subsubsection{Slow Rabi oscillations}
 In the case of small Jaynes-Cummings coupling, i.e. $g\ll\nu,\chi$, we focus on the parameter region $\beta\approx 1$ for strong optomechanical coupling and restrict the following discussion to the subspace of only one excitation in the quantum electrodynamical subsystem, remarking that the generalization is straightforward. In the interaction picture and after displacing the Hamiltonian~\eqref{eq.hamiltonian1} with $D(-\beta\vert g,1\rangle\!\langle g,1\vert)$ it takes on the form
\begin{align}
\label{eq:rwa1}
\tilde{H}=\hbar g\left[e^{i\beta^2\nu t} D(\beta e^{i\nu t}) \vert e,0\rangle\!\langle g,1\vert + {\rm H. c.}\right]\,,
\end{align}
apart from a constant energy, in the relevant subspace spanned by the two states $\vert g,1\rangle=\vert g\rangle\vert 1\rangle$ and $\vert e,0\rangle=\vert e\rangle\vert 0\rangle$. A rotating-wave approximation can be performed after expanding the displacement operator in a power series and taking only those terms that rotate with the smallest occurring frequency, which for $\beta\approx 1$ is given by $(\beta^2-1)\nu$. In the expansion of $D(\beta\exp[i\nu t])$ these surviving terms contain a single operator $b$ more than $b^\dagger$. Applying this procedure to~\eqref{eq:rwa1} leads to the approximate Hamiltonian
\begin{align}
\label{eq:rwa2}
\tilde{H}_{\rm RWA}=\hbar g\left[e^{i(\beta^2-1)\nu t}f(b^\dagger b)b\vert e,0\rangle\!\langle g,1\vert+ {\rm H. c.}\right]
\end{align}
where the function $f(m)$ can be obtained by evaluating the matrix elements $_{\rm mec}\langle m\vert D(\beta)\vert m+1\rangle_{\rm mec}$ \cite{cahill1969} which explicitly results in the expression
\begin{align}
	f(m)=\frac{-\beta}{m+1}e^{-\beta^2/2}L_m^{(1)}(\beta^2)
\end{align}
with the generalized Laguerre polynomials $L_m^{(\alpha)}(x)$. Within this approximation the Hamiltonian is a Jaynes-Cummings-type interaction between the atom-cavity subsystem and the mechanical oscillator with an additional phonon number dependent factor that does not alter the eigenstates, i.e. the dressed states. In the original picture the Hamiltonian then reads
\begin{align}
H_{\rm RWA}=&\hbar\nu b^\dagger b-\hbar\chi\vert g,1\rangle\!\langle g,1\vert(b+b^\dagger)\nonumber\\
&+\hbar g \left[ f(b^\dagger b)bD^\dagger(\beta)\vert e,0\rangle\!\langle g,1\vert+ {\rm H. c.}\right]
\end{align}
and its eigenstates are dressed states in which the state $\vert g,1\rangle\vert m\rangle_{\rm mec}$ is displaced by $D(\beta)$ while the state $\vert e,0\rangle\vert m+1\rangle_{\rm mec}$ remains undisplaced. For the initial state where the atom is prepared in its excited state and the mechanical oscillator in its ground state, i.e.  $\vert\psi(t=0)\rangle=\vert e,0\rangle\vert 0\rangle_{\rm mec}$, the time evolution results in the reduced density operators 
\begin{align}
\mu_{\rm RWA}(t)=&\left[\cos^2(\Omega t)+\cos^2\vartheta\sin^2(\Omega t)\right]\vert 0\rangle_{\rm mec}\langle 0\vert \nonumber\\
&+\sin^2\vartheta\sin^2(\Omega t) D(\beta)\vert 1\rangle_{\rm mec}\langle 1\vert D^\dagger(\beta)
\end{align}
and
\begin{align}
\rho_{\rm RWA}(t)=&\left[\cos^2(\Omega t)+\cos^2\vartheta\sin^2(\Omega t)\right]\vert e,0\rangle\!\langle e,0\vert\nonumber\\
&+\sin^2\vartheta\sin^2(\Omega t) \vert g,1\rangle\!\langle g,1\vert
\end{align}
for the mechanical oscillator and the atom-cavity subsystem, respectively. Here we defined $\tan\vartheta=\Omega/(\beta^2-1)\nu$ and $\Omega=2gf(0)$. We compared these approximate results with numerical propagation and found especially good agreement for the case $\beta=1$. From these outcomes we conclude: {\it (i)} The motion of the mechanical oscillator leads to non-classical states being composed of two contributions, namely an incoherent superposition of the vacuum and a displaced Fock state. They correspond to the radiation pressure of zero and one photon which are associated with the two trajectories $n=0$ and $n=1$ of Fig.~\ref{fig.trajectories}a) in the uncoupled case and become mixed due to the coherent Rabi oscillations between $\vert g,1\rangle$ and $\vert e,0\rangle$. {\it (ii)} For $\beta\neq 1$ one finds detuned Rabi oscillations of the atom-cavity subsystem, while they are strictly sinusoidal for $\beta=1$. {\it (iii)} The regime considered here is not sufficient for observing a suppression of Rabi oscillations of the atomic population. Nevertheless, it provides a first insight into the coupled dynamics of the system and the origin of the emergence of non-classicalities in the state of the mechanical oscillator.

\subsubsection{Small optomechanical coupling}
\label{sec:limiting2}
Another limit which allows for an analytical treatment is the regime of small values of $\beta$ where Ref.~\cite{restrepo2014} provides an approximation of the Hamiltonian~\eqref{eq.hamiltonian1}. According to that treatment, in a frame displaced by $D(-\beta/2)$ with $\beta\ll 1$ and apart from a constant term, the dynamics is governed by the Hamiltonian
\begin{align}
	\label{eq:perturbative1}
	\tilde{H}_\beta=&\hbar\nu b^\dagger b+\hbar g\left[\vert +\rangle\!\langle +\vert-\vert -\rangle\!\langle -\vert\right]\nonumber\\
	&-\frac{\beta}{2}\hbar\nu\left[b \vert +\rangle\!\langle -\vert +{\rm H.c.} \right]\,, 
\end{align}
in the subspace spanned by the usual Jaynes-Cummings dressed states $\ket{\pm}=[\ket{g,1}\pm\ket{e,0}]/\sqrt{2}$ of the resonant atom-cavity subsystem. Eq.~\eqref{eq:perturbative1} is again of Jaynes-Cummings form, with the ladder operators $\vert \pm\rangle\!\langle \mp\vert$ playing the role of the atomic rising and lowering operators and the mechanical oscillator taking over the cavity part. The approximate form~\eqref{eq:perturbative1} is based on a rotating wave approximation which additionally requires $|2g-\nu|\ll 2g+\nu$. In the case of a resonant interaction, $2g=\nu$, the double dressed states
\begin{align}
\vert 0,0\rangle=&\vert -\rangle\vert 0\rangle_{\rm mec}\quad \text{(for $m=0$)}\,,\\
\vert \pm,m\rangle=&\frac{1}{\sqrt{2}}\left[\vert +\rangle \vert m-1\rangle_{\rm mec}
\pm\vert -\rangle \vert m\rangle_{\rm mec}\right]
\end{align}
are the eigenstates of the Hamiltonian~\eqref{eq:perturbative1} with the corresponding eigenfrequencies $\varepsilon_0=-\nu/2$ and
\begin{align}
 \varepsilon_m^{(\pm)}=\left[m-\frac{1}{2}\mp\frac{\beta}{2}\sqrt{m}\right]\nu
\end{align}
for $m\ge1$. 
The initial state $\ket{\psi(t=0)}=\ket{e,0}\ket{0}_\text{mec}$ can be propagated with help of the Hamiltonian~\eqref{eq:perturbative1} and leads to
\begin{widetext}
\begin{align}
\ket{\tilde\psi(t)}=e^{-\frac{\beta^2}{8}}\sum_{m=0}^\infty\frac{\left(-\frac{\beta}{2}e^{-i\nu t}\right)^m}{\sqrt{2m!}}\Bigg\{&-i\sin\left(\sqrt{m}\frac\beta{2}\nu t\right)e^{i\frac{\nu}{2}t}\ket{+}\ket{m-1}_{\rm mec}\nonumber\\
&+\left[\cos\left(\sqrt{m+1}\frac\beta{2}\nu t\right)e^{-i\frac{\nu}{2} t}\ket{+}-\cos\left(\sqrt{m}\frac\beta{2}\nu t\right)e^{i\frac{\nu}{2}t}\ket{-}\right]\ket{m}_{\rm mec}\nonumber\\
&+i\sin\left(\sqrt{m+1}\frac\beta{2}\nu t\right)e^{-i\frac{\nu}{2} t}\ket{-}\ket{m+1}_{\rm mec}\Bigg\}
\end{align}
\end{widetext}
in the displaced frame. More insight reveals the $m=0$ term,
\begin{align}
\label{eq:approx_state}
\ket{\psi(t)}\!\approx\! \frac{e^{-\frac{\beta^2}{8}}}{\sqrt{2}}&
\left\{\!\left[
\cos\left(\frac\beta{2}\nu t\right)\!e^{-i\frac\nu{2}t}\ket{+}\!-\!e^{i\frac\nu{2}t}
\ket{-}\right]\!\ket{\beta/2}_{\rm mec}\right.\nonumber\\
&\!\!\!\!\!\left.+i\sin\left(\frac\beta{2}\nu t\right)e^{-i\frac\nu{2}t}\ket{-}D\big(\tfrac\beta{2}\big)\ket{1}_{\rm mec}\right\}\,,
\end{align}
written here in the original frame, showing the dominant behaviour of $\ket{\psi(t)}$ for $\beta\ll 1$: The first term describes the usual Rabi-oscillations of the atom-cavity system when $\beta\to 0$, whereas the second term represents an oscillating displaced number state of the mechanical oscillator, associated with the dressed state $\ket{-}$. Nevertheless, the condition $\beta\ll 1$ for the validity of the Hamiltonian~\eqref{eq:perturbative1} prevents clear signatures of non-classicality of the evolved state in this limit.

The time evolution of the excited state population $P_e(t)={\rm Tr}[\vert e\rangle\!\langle e \vert\psi(t)\rangle\!\langle\psi(t)\vert]$, using the state~\eqref{eq:approx_state}, can be further approximated by
\begin{align}
P_e(t)\approx \frac{1}{2}\left[1+\cos\left(\frac{\beta}{2}\nu t \right)\cos(\nu t)\right]
\end{align}
which constitutes a sinusoidally modulated Rabi oscillation with a beat frequency of $\beta\nu/2$ which describes the onset of the suppression of Rabi-oscillations for small values of $\beta$.

\subsection{Suppression of Rabi oscillations}
We now focus on the strong-coupling regime $\beta\gtrsim 1$ where the approximate Hamiltonians are not valid and one has to deal with the hybrid Hamiltonian, Eq.~\eqref{eq.hamiltonian1}, that can be mapped onto a driven Rabi model~\cite{rabi1936}. Since such a model does not provide easy-to-handle analytic solutions~\cite{braak2011,zhong2013} we restrict ourselves in the following on the discussion of numerical results. 
\subsubsection{The case $g=\nu$}
\begin{figure}[h!]
	\flushleft{a)}\vspace{-4ex}
	\begin{center}
		\includegraphics[width=0.9\linewidth]{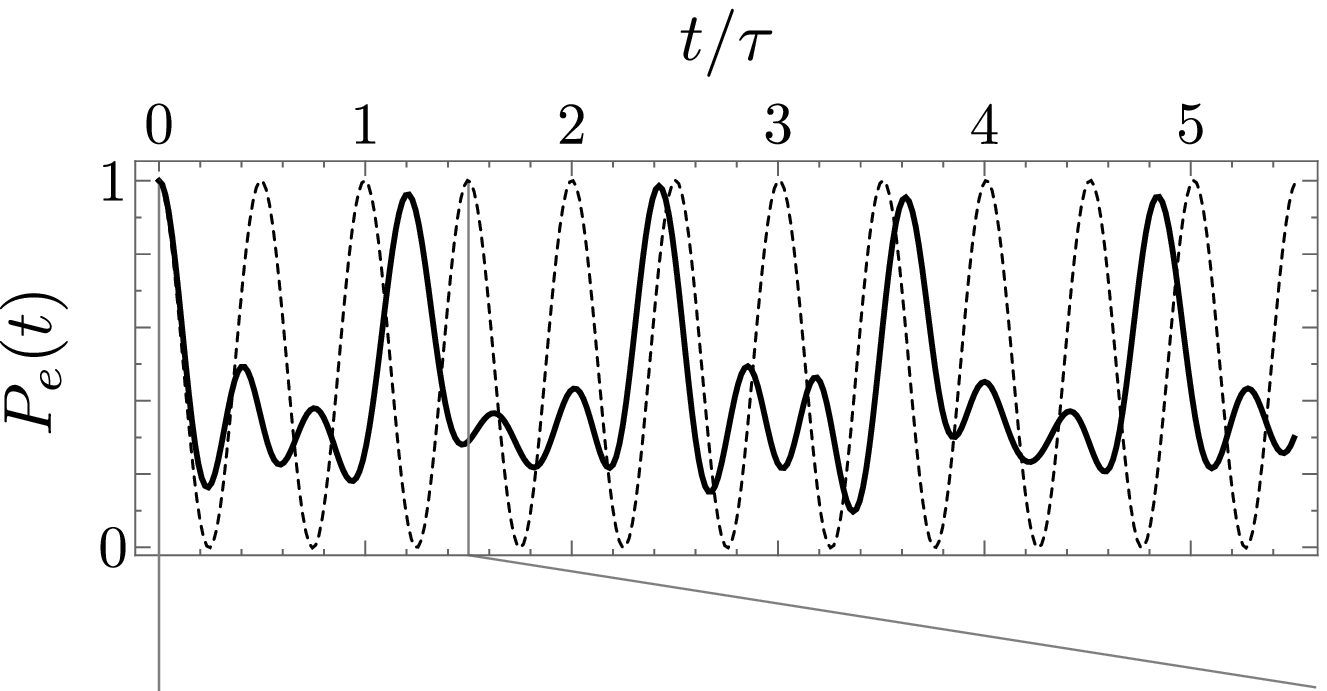}
	\end{center}\vspace{-6.7ex}
	\flushleft{b)}\vspace{-4ex}
	\begin{center}
		\includegraphics[width=0.9\linewidth]{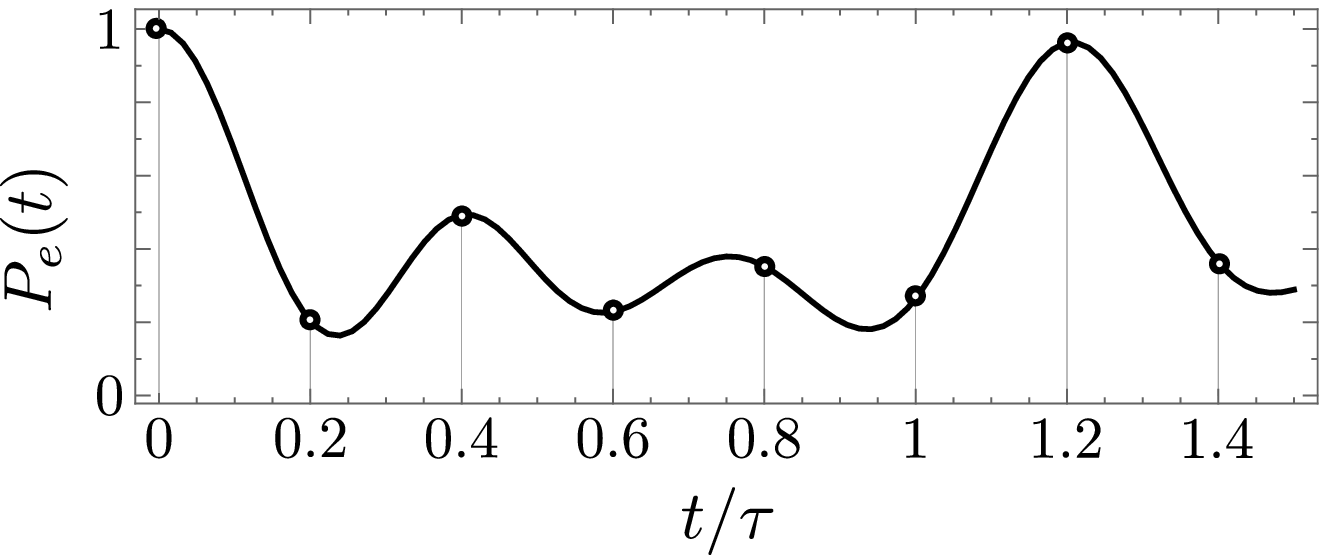}
	\end{center}
	\caption{\label{fig.blockade} a) Time evolution of the excited state population for the initial state $\vert e,0\rangle\vert 0\rangle_{\rm mec}$. The population experiences a strong suppression before a first revival at a time larger then a Rabi period. The parameters are $\beta=1$ and $g=\nu$. For comparison the unperturbed Rabi oscillations ($\beta=0$) are shown in the dashed line. b) Inset on a time scale until the first revival occurs. The circles indicate instances of time which will be referred to later.}
\end{figure}

We first focus on the case $g=\nu$ and $\beta=1$.  We start again from the same initial state $\vert\psi(t=0)\rangle=\vert e,0\rangle\vert 0\rangle_{\rm mec}$ and propagate it numerically using the Hamiltonian~\eqref{eq.hamiltonian1}. In Fig.~\ref{fig.blockade} we show the time evolution of the excited state population $P_e(t)={\rm Tr}[\vert e\rangle\!\langle e \vert\psi(t)\rangle\!\langle\psi(t)\vert]$.
The strictly periodic Rabi oscillation of the uncoupled case (dashed) undergoes a drastic change. Instead of the sinusoidal time evolution the excited state population exhibits a strong suppression before it rises up again at a time appreciably longer than the Rabi period $\pi/g$. This behavior continues quasi-periodically for the considered parameters and can be qualitatively understood with the following intuitive explanation: As time evolves the atom initially prepared in the excited state populates the cavity with a single photon leading to a rising radiation pressure inside the cavity pushing the mirror significantly out of its equilibrium position and driving it along the trajectory $n=1$ in Fig.~\ref{fig.trajectories}a). Its displacement is accompanied by an effective dynamical detuning between atom and cavity suppressing the Rabi oscillations. After approximately one mechanical oscillator period $\tau=2\pi/\nu$ the atom-cavity resonance condition is fulfilled again and the Rabi oscillation continues. This simple picture however does not explain all details in Fig.~\ref{fig.blockade}: The behavior is not strictly periodic in the mechanical period $\tau$ and $P_e(t)$ does not show a full revival. 

\begin{figure}[h!]
\flushleft{a)}\vspace{-4ex}
\begin{center}
\includegraphics[width=0.9\linewidth]{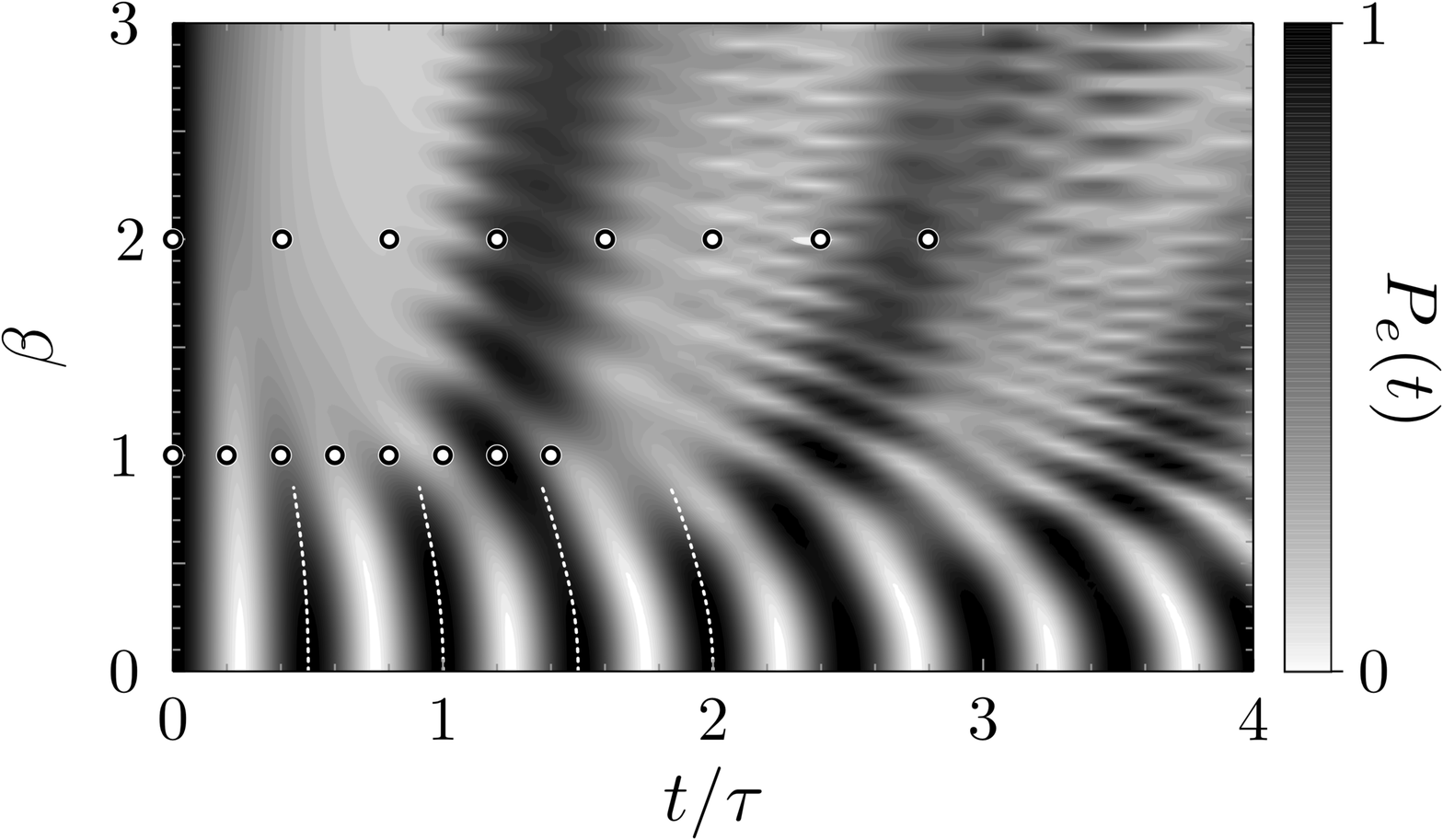}
\end{center}\vspace{-5ex}
\flushleft{b) $\beta=1$}\vspace{-1ex}
\begin{center}
\includegraphics[width=0.9\linewidth]{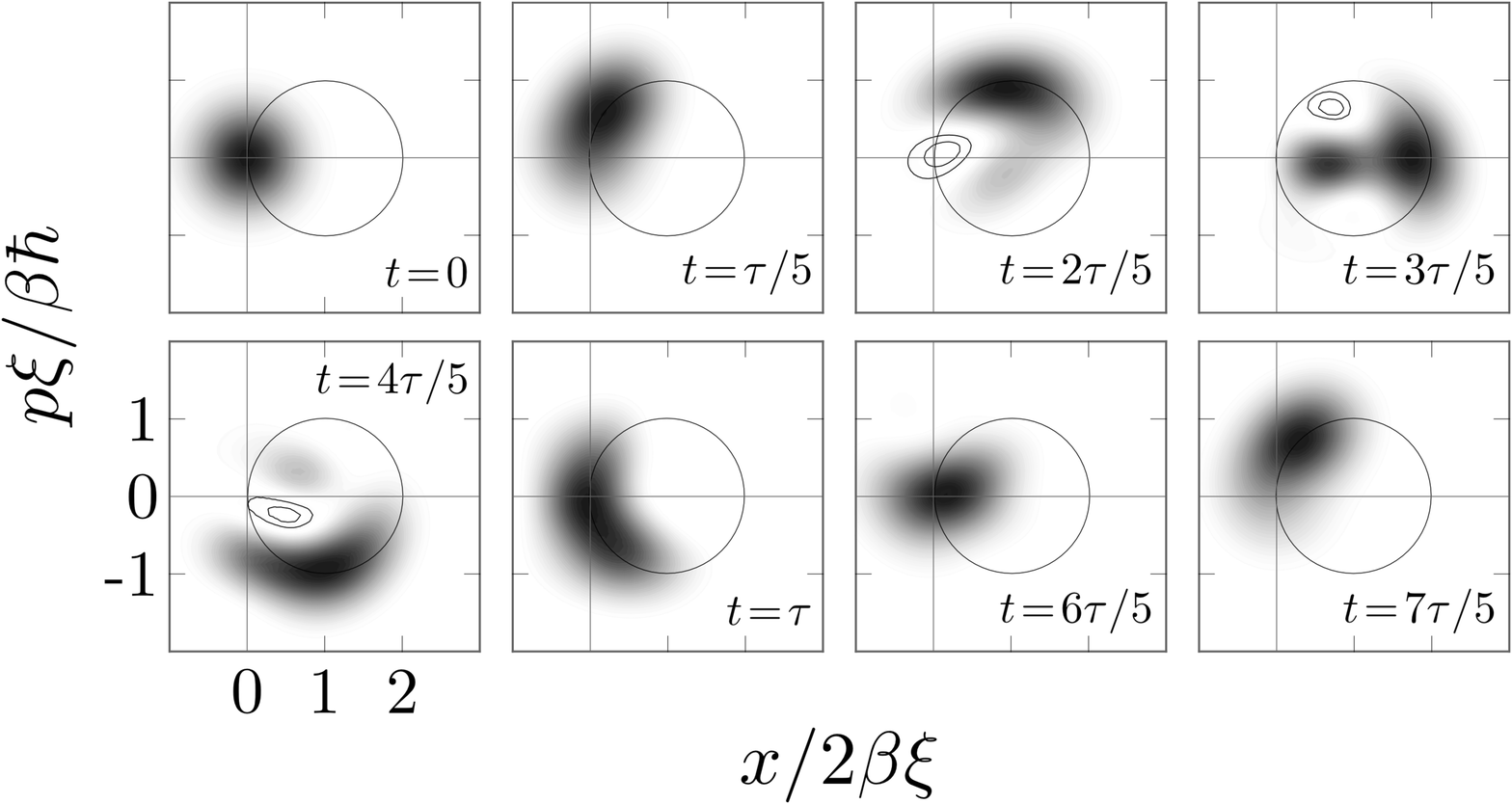}
\end{center}\vspace{-5ex}
\flushleft{c) $\beta=2$}\vspace{-1ex}
\begin{center}
\includegraphics[width=0.9\linewidth]{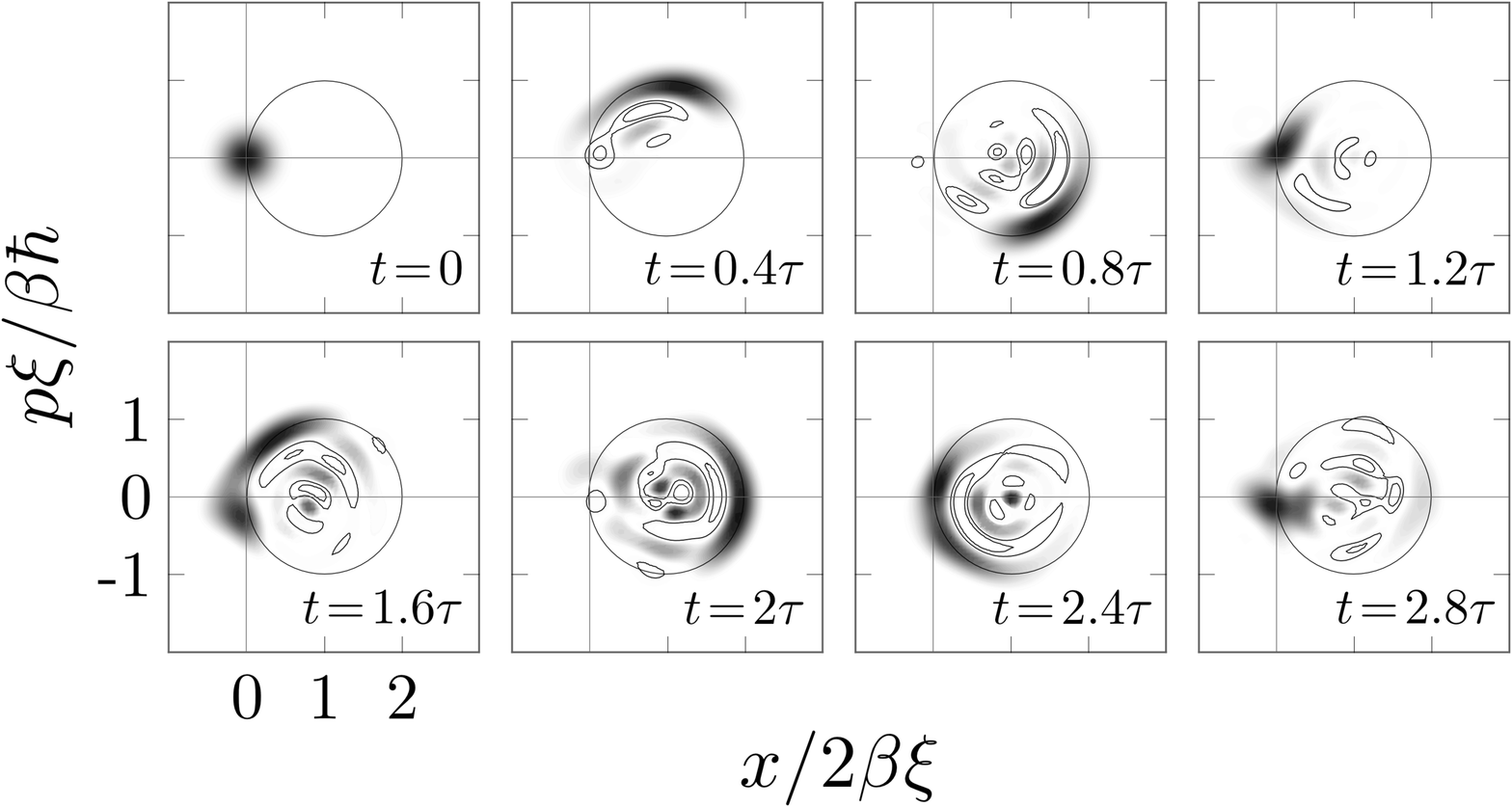}
\end{center}
\caption{\label{fig.blockade_contour}a) Dependence of the Rabi oscillations on the optomechanical coupling ratio $\beta$ for $g=\nu$. The dotted lines show the dependence of the times at which the population exhibits maxima in dependence of $\beta$ following from a perturbative treatment of the optomechanical coupling. b) Wigner function of the mechanical oscillator for the eight instances denoted by the circles in a) along the line $\beta=1$. c) Wigner function of the mechanical oscillator for the eight instances denoted by the circles in a) along the line $\beta=2$. Encircled areas denote regions where the Wigner function takes on negative values.}
\end{figure}

To systematically analyze the suppression of the Rabi oscillations for different values of $\beta$ using the same initial state $\vert\psi(t=0)\rangle=\vert e,0\rangle\vert 0\rangle_{\rm mec}$ as before we show in Figure~\ref{fig.blockade_contour}a) the time evolution of the population of the excited state in dependence of $\beta$ color-coded in a density plot for $g=\nu$ where dark shading indicates high values of the population in $\vert e\rangle$.
For the dotted white lines in the lower part of Fig.~\ref{fig.blockade_contour}a) we used the perturbative results of Sec.~\ref{sec:limiting2} to show the times $T_n$ of the $n$-th maximum of the perturbed Rabi oscillation  for the borderline parameters chosen here. These times can be approximated by the expression $T_n\approx n\pi[1-\widetilde{m}_n\beta^2/2]/g$, where the $\widetilde{m}_n$ play the role of average phonon numbers. 
As $\beta$ increases this quadratic behaviour ceases and distinct maxima of the population emerge separated by gaps approximately given by the mechanical period $\tau$ being significantly longer than a Rabi period $\tau/2$ for the parameters. For even larger $\beta\lesssim 3$ the first revival stabilizes while subsequent revivals of the population show interference-like patterns and decline.

In order to provide deeper insight into this behaviour we show in Figs.~\ref{fig.blockade_contour}b) and c) the Wigner function of the mechanical oscillator for two different values of $\beta$ and eight different instances of time as marked by the open circles in the density plot Fig.~\ref{fig.blockade_contour}a) (and in Fig.~\ref{fig.blockade}b) for $\beta=1$). For both cases the mechanical oscillator starts in the vacuum state represented by the Gaussian Wigner function of minimal uncertainty. For $\beta=1$ the phase space distribution begins to follow the trajectory belonging to the radiation pressure of a single photon inside the cavity until at $t\approx2\tau/5$ the quasi probability distribution splits into two contributions and the non-classicality of this state manifests itself in the negative parts in between which are denoted by encircled areas. This behavior is caused by the coherent swapping between the states $\vert e,0\rangle$ and $\vert g,1\rangle$ of different radiation pressures corresponding to the two trajectories $n=0$ and $n=1$ in Fig.~\ref{fig.trajectories}a). At $t\approx3\tau/5$ the two contribution are clearly distinguishable and merge again towards the end of the mechanical period at $t\approx\tau$ hence bringing atom and cavity back into resonance. One can observe however that the Wigner function is mostly localized around $t\approx6\tau/5$, i.e. after more than one mechanical period. We conjecture that this delay stems from the Rabi oscillations taking place simultaneously with the oscillation of the mechanical element. After that, the quasi-periodic behavior starts into another round. For $\beta=2$ the time evolution of the Wigner function is coined by the appearance of strong interferences in phase space for almost all instances of time as shown in Fig.~\ref{fig.blockade_contour}c). Only at $t=0$ and the first revival at $t\approx 1.3\tau$ it is positive, already at the next revival at $t\approx 2.6\tau$  negativities  and a delocalized Wigner function can be observed. This spread in phase space and the interferences explain incomplete population of the excited state and the fading of the later revivals: no sharp mirror position and thereby no defined resonance between atom and cavity can be found for later times. We also checked if the blockade effect persists for higher photon numbers and found indeed very similar characteristics.

\subsubsection{The case $g=\nu/2$}
\begin{figure}[h!]
\flushleft{a)}\vspace{-4ex}
\begin{center}
\includegraphics[width=0.9\linewidth]{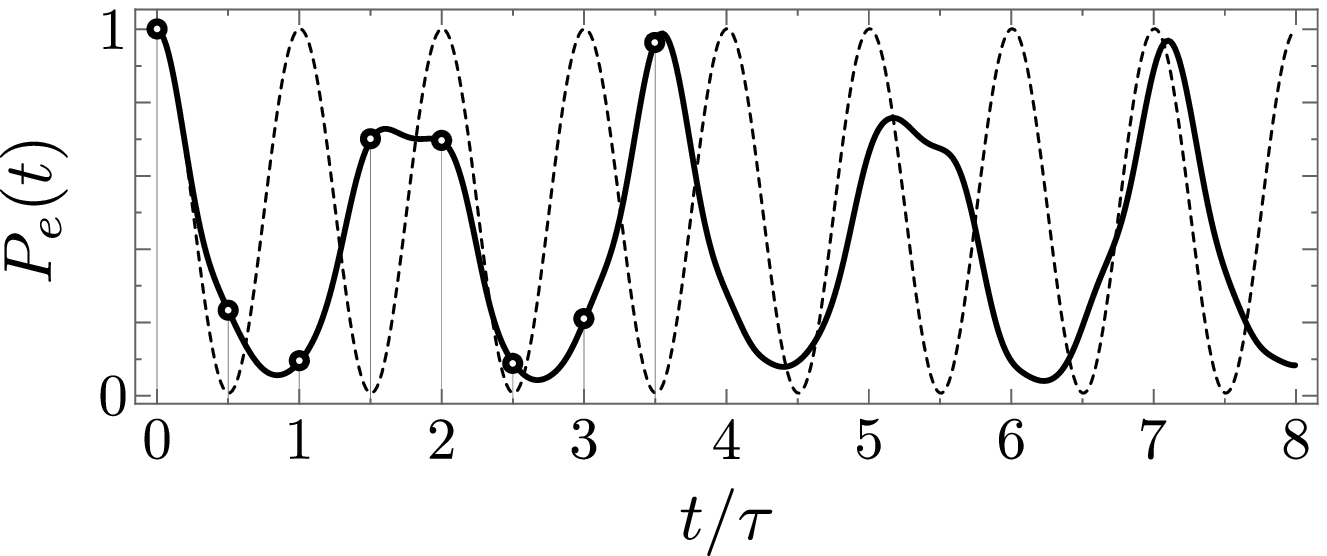}
\end{center}\vspace{-6.7ex}
\flushleft{b)}\vspace{-4ex}
\begin{center}
\includegraphics[width=0.9\linewidth]{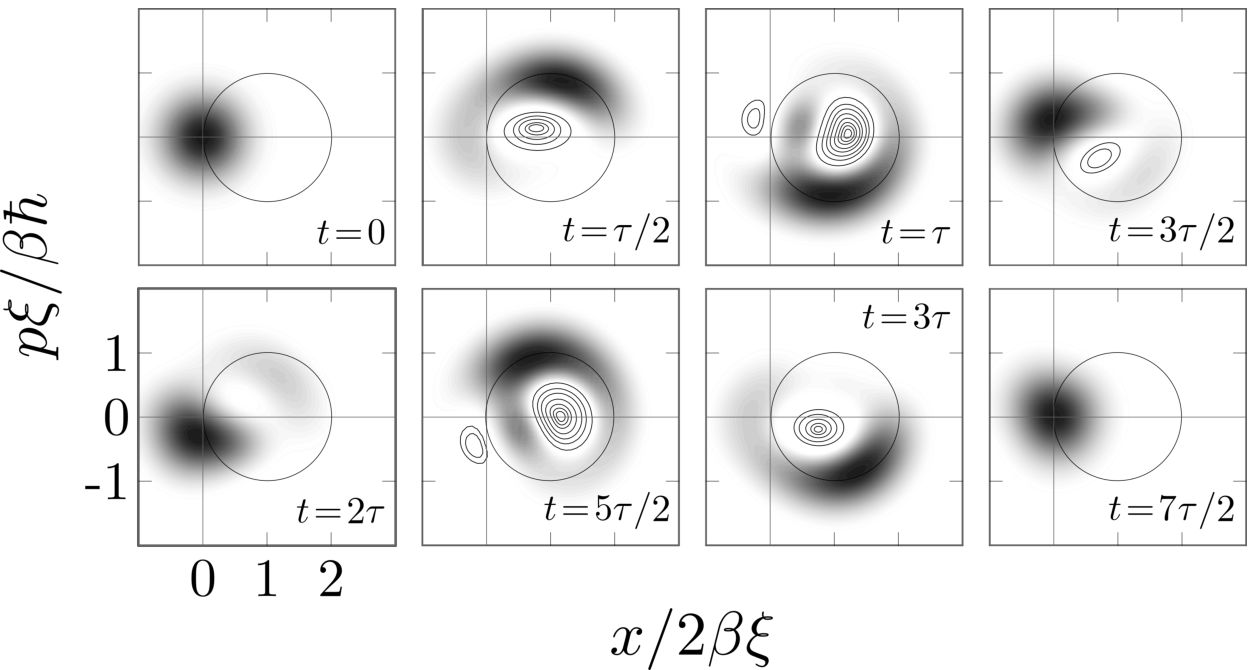}
\end{center}
\caption{\label{fig.blockade2} a) Time evolution of the excited state population for the initial state $\vert e,0\rangle\vert 0\rangle_{\rm mec}$ for $\beta=1$ and $g=\nu/2$. The unperturbed Rabi oscillations ($\beta=0$) are shown in the dashed line. b) Wigner function of the mechanical oscillator state at the eight instances denoted by open circles in a). The encircled areas indicate regions where the Wigner functions takes on negative values, more circles correspond to more negative values.}
\end{figure}

We briefly consider the case $g=\nu/2$, whose limit for small $\beta$ was analyzed in Sec.~\ref{sec:limiting2}. The numerical results in this case for $\beta=1$ are presented in Fig.~\ref{fig.blockade2}.
The excited state population in subfigure a) alternatingly displays incomplete and almost full revivals separated by roughly $\Delta t=1.75\tau$.  In contrast to the previous case $g=\nu$, the Wigner functions of the mechanical oscillator shown in Fig.~\ref{fig.blockade2}b) exhibit much more pronounced negativities. This phase space structure of the mechanical oscillator can be interpreted as an extrapolation of the simple superposition of vacuum and Fock state $\ket{1}_{\rm mec}$, Eq.~\eqref{eq:approx_state}, being present for $\beta\to 0$, towards $\beta=1$.

Before we go over to the dissipative dynamics we point out the connection of our model to the photon blockade in optomechanical systems. The photon blockade~\cite{rabl2011} results from the interaction between the cavity field mode and the mechanical object due to the radiation pressure which lifts the degeneracy of the equidistant spectrum of the harmonic oscillators. Descriptively expressed, if a resonant photon is scattered into the cavity, a second photon is effectively detuned from resonance and therefore reflected with increased probability. When the entering photon stems from a second cavity which coherently couples to the optomechanical resonator in the form of a beam-splitter-like interaction, the analogy to the model presented is established by replacing this second cavity with a two-level atom.

\section{Dissipative dynamics}
In this section we include spontaneous decay of the atom with rate $\Gamma$, cavity losses  at rate $\kappa$ and damping of the mechanical oscillator on a time scale $1/\gamma$ into account. In the regime considered in this work the dissipative dynamics can be described by the master equation
\begin{align}
\label{eq.masterequation}
\frac{\partial\varrho}{\partial t}=\mathcal{L}\varrho= \frac{1}{i\hbar}[H,\varrho]+\mathcal{L}_{\rm tls}\varrho+ \mathcal{L}_{\rm cav}\varrho
+ \mathcal{L}_{\rm mec}\varrho
\end{align}
in Born-Markov approximation. The strong optomechanical coupling requires a treatment where the damping of the cavity and the mechanical oscillator cannot be treated independently~\cite{ponte2004,hu2015}. The non-unitary parts of the dynamics are given by
\begin{align}
\mathcal{L}_{\rm tls}\varrho&=\frac{\Gamma}{2}\mathcal{D}[\vert g\rangle\!\langle e\vert]\varrho\,,\\
\mathcal{L}_{\rm cav}\varrho&=\frac{\kappa}{2}\mathcal{D}[a]\varrho+4\gamma\frac{k_{\rm B}T}{\hbar\nu}\mathcal{D}[\beta a^\dagger a]\varrho\,,  \\
\mathcal{L}_{\rm mec}\varrho&=\frac{\gamma}{2}\bar{m}\mathcal{D}[b^{\dagger}-\beta a^\dagger a]\varrho + \frac{\gamma}{2}\big(\bar{m}+1\big)\mathcal{D}[b-\beta a^\dagger a]\varrho
\end{align}
where we introduced the super-operator notation $\mathcal{D}[X]\varrho=2X\varrho X^\dagger-X^\dagger X\varrho-\varrho X^\dagger X$ to describe Lindblad terms. The mean thermal phononic occupation at temperature $T$ is given by $ \bar{m}=[\exp(\hbar\nu/k_{\rm B}T)-1]^{-1}$. Again the same initial state $\varrho(t=0)=\vert\psi(t=0)\rangle\!\langle\psi(t=0)\vert$ was chosen and its time evolution was calculated by numerical diagonalization of the Liouville operator $\mathcal{L}$ in Eq.~\eqref{eq.masterequation}.

\begin{figure}[h!]
\flushleft{a)}\vspace{-4ex}
\begin{center}
\includegraphics[width=0.9\linewidth]{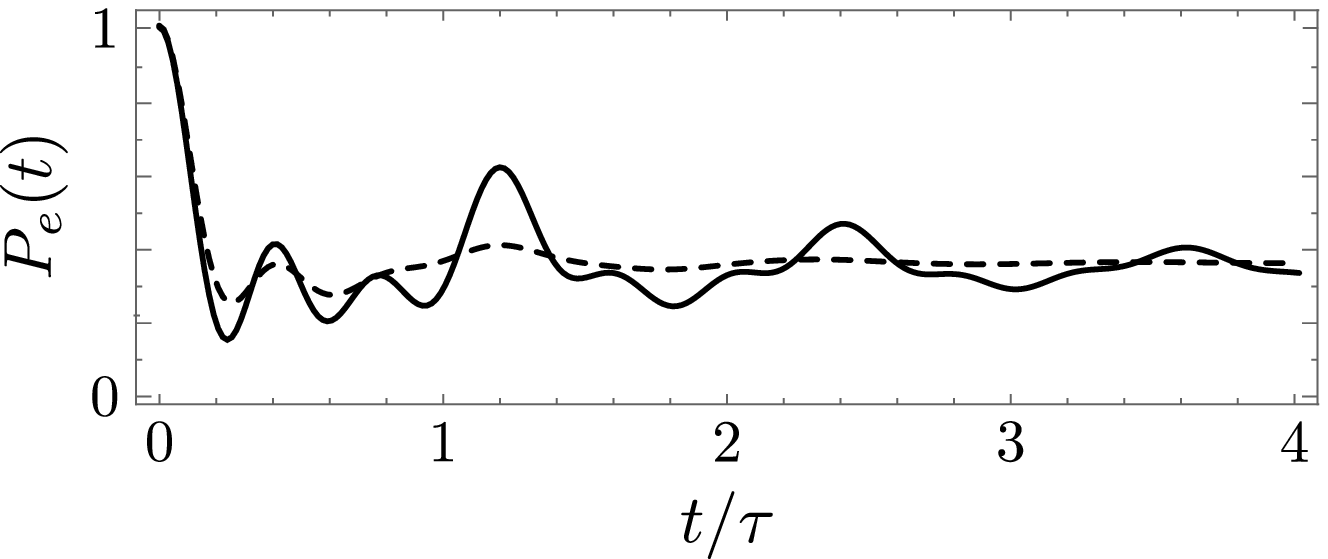}
\end{center}\vspace{-4ex}
\flushleft{b)}\vspace{-4ex}
\begin{center}
\includegraphics[width=0.9\linewidth]{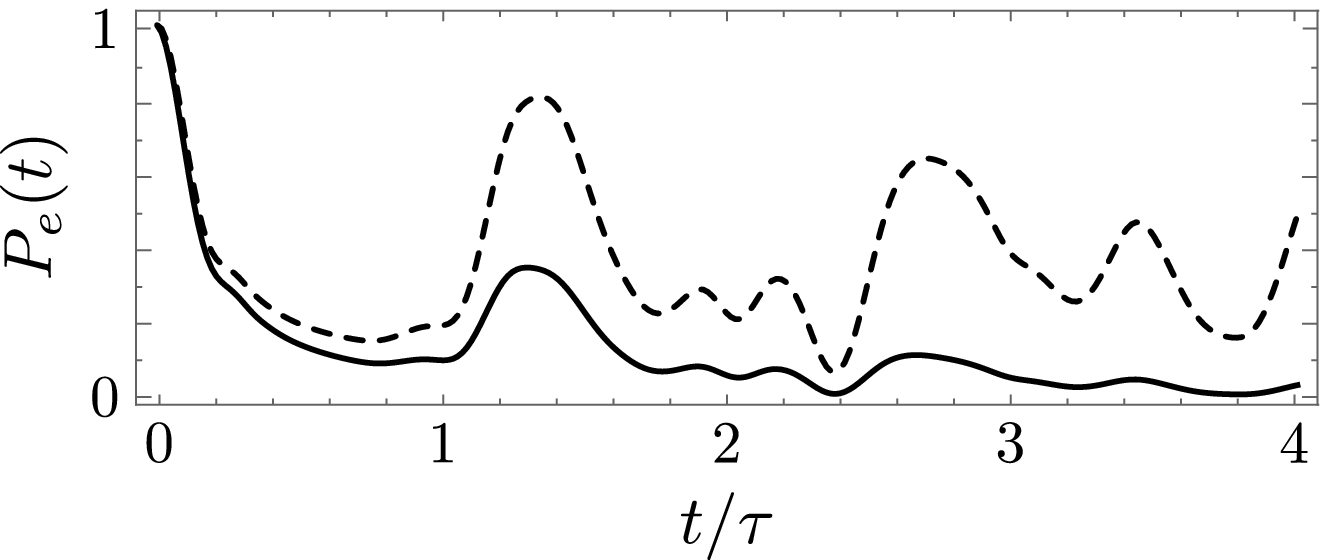}
\end{center}\vspace{-4ex}
\flushleft{c)}\vspace{-4ex}
\begin{center}
\includegraphics[width=0.9\linewidth]{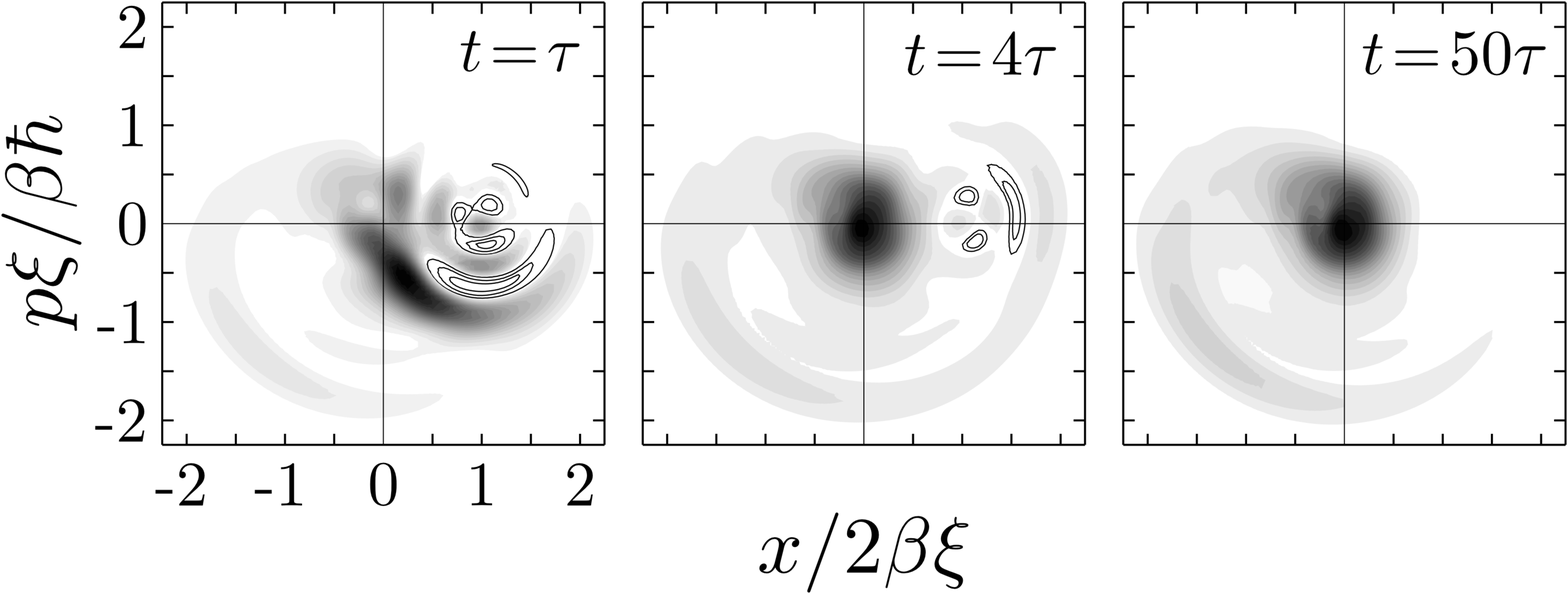}
\end{center}
\caption{\label{fig.blockade-dissipation} Time evolution of the excited state population under dissipative dynamics for $g=\nu$. a) The bad oscillator case ($\Gamma=\kappa=0$) for $\beta=1$. The mean phononic occupation is $\bar{m}=0.25$ and the mechanical decay rates are $\gamma=0.1\nu$ (solid) and $\gamma=0.4\nu$ (dashed). b) The good oscillator case ($\gamma$=0) with the atomic and cavity decay rates $\Gamma=\kappa=0.1\nu$ (solid) and the undamped case (dashed) for $\beta=2$. c) Wigner function of the mechanical oscillator for the good resonator case of b) at times $t=\tau$, $t=4\tau$ and $t=50\tau$. Encircled areas denote negativities of the Wigner function.}
\end{figure}

When the dissipative dynamics is dominated by the mechanical damping, i.e. $\gamma\gg \kappa,\Gamma$, the mechanical oscillator approaches its thermal state and thereby averages out the Rabi oscillations of the atom-cavity subsystem towards a constant value of the atomic population as shown in Fig.~\ref{fig.blockade-dissipation}a) for $\gamma=0.1\nu$ (solid) and $\gamma=0.4\nu$ (dashed). Nevertheless, the suppressed Rabi oscillation are observable even for such mechanical decay rates which are much larger than in typical existing or planned optomechanical setups with $Q$-factors in the order of $10^3-10^6$~\cite{cohen2014,wilson2015,kipfstuhl2014}.

For high-$Q$ mechanical oscillators the relevant time scale for damping is given by the atomic and cavity decay. Such a situation is depicted in Fig.~\ref{fig.blockade-dissipation}b) for $\kappa=\Gamma=0.1\nu$ and $\beta=2$ (solid) where we also show the undamped case (dashed) for comparison. For such high values of $\beta$ where the Wigner function exhibits pronounced negativities one might suppose that the virtually undamped mechanical oscillator is capable to support a non-classical quasi-stationary state. However, the snapshots in Fig.~\ref{fig.blockade-dissipation}c) reveal that the time evolution disembogues into a non-trivial quasi-stationary Wigner function where however all negativities are averaged out due to the randomness of the spontaneous emission and cavity losses. We remark that in an experiment where the emitted photon is recorded the negativity of the Wigner function persist for such a single quantum trajectory~\cite{plenio1998}.

\section{Initial states}
We eventually analyze the dependence of the modified Rabi oscillations on the initial state. Current technology provides techniques which allow for the preparation of atomic and cavity states to a high degree of fidelity. The initial state of the mechanical oscillator is prepared by laser cooling. Most relevant initial states to be considered are hence displaced thermal states of the form
\begin{align}
\mu_{\alpha,\bar{m}}=D(\alpha)\left[\frac{1}{\bar{m}+1}\left(\frac{\bar{m}}{\bar{m}+1}\right)^{b^\dagger b}\right]D^\dagger(\alpha)\,,
\end{align}
such that $\varrho(t=0)=\vert e,0\rangle\!\langle e,0\vert\mu_{\alpha,\bar{m}}$. The initial mechanical displacement is characterized by the complex value $\alpha$ and the extension in phase space is proportional to the mean thermal occupation $\bar{m}$. The variance of the mechanical number operator in these states has the value $\Delta m^2=\bar{m}(\bar{m}+1)+(2\bar{m}+1)|\alpha|^2$. We restrict ourself to the familiar case of $g=\nu$ and $\beta=1$. In order to compare the behavior of the modified Rabi oscillations with the ideal case, i.e. an initial mechanical vacuum, we adopt the measure
\begin{align}
\label{eq.measure}
F(\alpha,\bar{m})=\int_0^{6\tau/5}{\rm d}t\, [P_e^{0,0}(t)-P_e^{\alpha,\bar{m}}(t)]^2\,,
\end{align}
representing the mean quadratic deviation of the atomic population $P_e^{\alpha,\bar{m}}(t)$, belonging to $\mu_{\alpha,\bar{m}}$, from the ideal case $P_e^{0,0}(t)$, integrated over a time window where the suppression takes place. 
  \begin{figure}[h!]
  	\flushleft{a)}\vspace{-4ex}
  	\begin{center}
  		\includegraphics[width=0.9\linewidth]{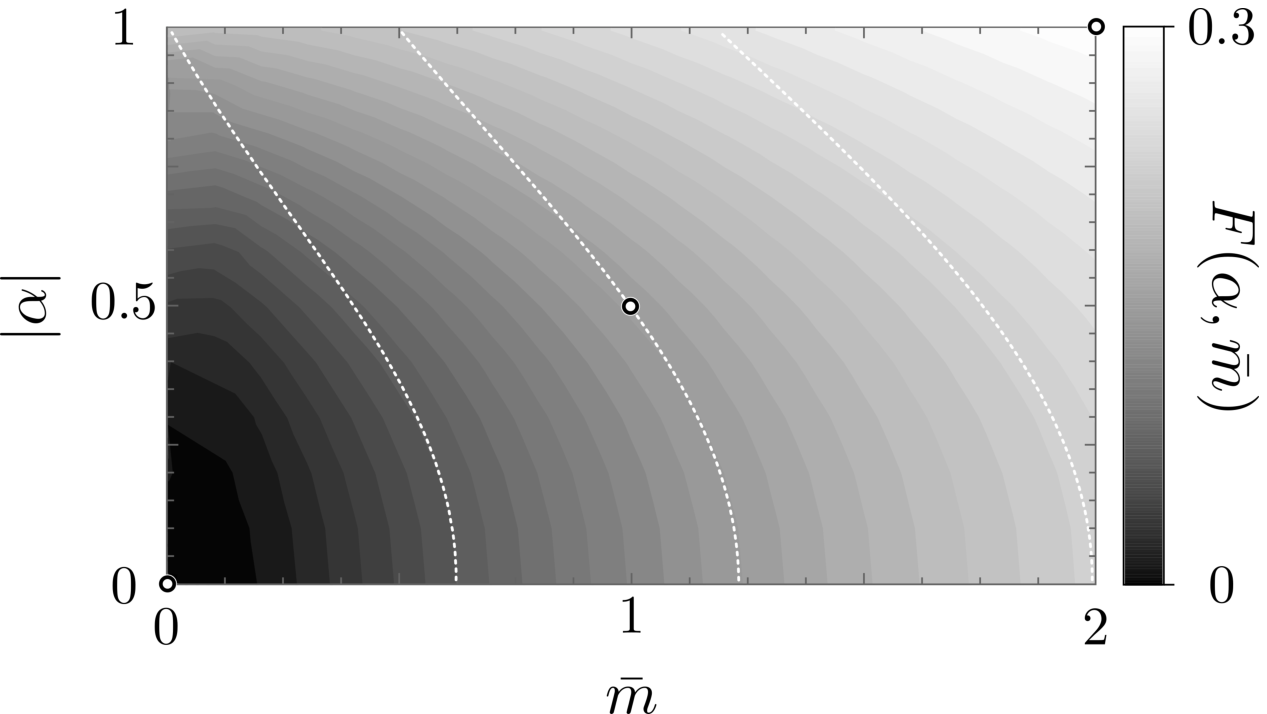}
  	\end{center}
  	\flushleft{b)}\vspace{-4ex}
  	\begin{center}
  		\includegraphics[width=0.9\linewidth]{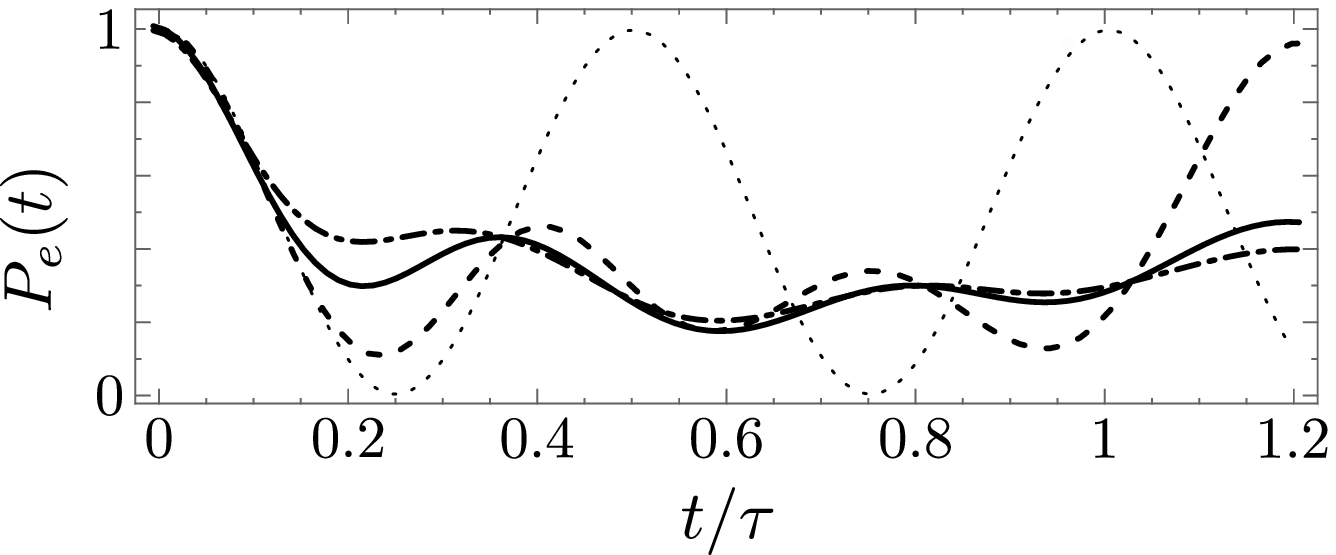}
  	\end{center}
  	\caption{\label{fig.initial}a) Density plot of the measure $F(\alpha,\bar{m})$ for the suppression of the Rabi oscillation in dependence of $|\alpha|$ and $\bar{m}$ of the initial state (for each modulus $|\alpha|$ an average over eight equidistant polar angles was performed). The dashed lines indicate the constant values of the phonon number variance $\Delta m^2=1,2.75,6$. b) Shows two exemplary evolutions of the excited state population for the initial states denoted by open circles in a) with $|\alpha|=0.5$, $\bar{m}=1$ (solid line) and $|\alpha|=1$, $\bar{m}=2$ (dashed-dotted line) corresponding to the values $F=0.15$ and $F=0.25$, respectively. As references the case of the usual initial state, viz. $\alpha=0$, $\bar{m}=0$, for $\kappa=\Gamma=0$ is shown in the dashed line and the unperturbed Rabi oscillations in the dotted line. The parameters are $g=\nu$, $\beta=1$, $\gamma=0$ and $\kappa=\Gamma=0.05\nu$.}
  \end{figure}	
In Fig~\ref{fig.initial}a) we show a density plot of $F(\alpha,\bar{m})$ for  $|\alpha|\leq 1$ and $\bar{m}\leq 2$ in the good resonator case with $\kappa=\Gamma=0.05\nu$ and coupling strengths $g=\nu$ and $\beta=1$. For the complex parameter $\alpha$ we preformed an average over eight equidistant polar angles for each modulus $|\alpha|$. From the density plot a quite stable behavior of the suppressed Rabi oscillations can be read off, which only marginally diminishes for temperatures corresponding to $\bar m\lesssim 0.25$ and displacements up to $|\alpha|=0.5$, but even higher temperatures and displacements are tolerable.

To clarify the expressiveness of the deployed measure we additionally depict the time-evolution of the atomic population in Fig~\ref{fig.initial}b) for the different initial states indicated by open circles in Fig.~\ref{fig.initial}a), i.e. $|\alpha|=0.5$, $\bar{m}=1$ and $\alpha=1$, $\bar{m}=2$ which correspond to $F=0.15$ and $F=0.25$, respectively. Strongly displaced thermal states hence do not exhibit an appreciable maximum in the region around $t=6\tau/5$, as found in the ideal case (dashed line and Fig.~\ref{fig.blockade}).
 The higher the displacement and mean thermal occupation the flatter the curves for the time evolution of the excited state population become, thereby drastically diminishing the visibility of the suppression effect. Nevertheless, within a quite large region in the parameter space of the initial states, the effect remains clearly distinguishable.

\section{Conclusion}
We identified a distinctive feature in the highly non-linear regime of a hybrid optomechanical system which manifests itself in the suppression of Rabi oscillations of the atom-cavity subsystem and dynamically produces non-classical Wigner functions in the mechanical degree of freedom. The presented phenomenon is linked to blockade effects of non-linear systems, particularly to the photon blockade of optomechanics. We analyzed the dynamics of this suppression and gave a qualitative and intuitive explanation using the time evolution of the mechanical Wigner function. We also pointed out that even under the influence of dissipation the main features of the effect are still observable. 

For the case of a resonant atom-cavity interaction, the effect is clearly observable when $\beta\gtrsim 1$ and $g\approx\nu$. Since it is based on the coherent time evolution, strong coupling is required, meaning $\kappa,\Gamma\ll g$ and $\gamma\ll\nu$. The latter requirement is already achieved in the majority of optomechanical experiments with high-$Q$ mechanical elements. Ground state cooling is required such that the initial state fulfills $\bar{m}\lesssim 1$. We exemplify these requirement based on current experiments, such as the experiment of Ref.~\cite{chan2011}, with a mechanical frequency $\nu=2\pi\cdot 4\,{\rm GHz}$, mechanical (optical) $Q$-factor of $Q=10^5$ ($Q=10^6$) and $\bar{m}\approx 1$, only lacking the necessary strong optomechanical coupling which should, along with the cavity-dipole coupling, also be in the GHz range. In upcoming hybrid-systems with diamond-based crystal cavities~\cite{riedrich-moller2012,kipfstuhl2014}, a resonant interaction with color centers with linewidths in the MHz range could be tailored, basically fulfilling the requirements for the observation of the suppression effect presented here.

With this work we provided deeper insight into the rather unexplored area of the non-linear dynamics of hybrid optomechanical systems being potentially pivotal for future quantum technological applications interfacing different quantum degrees of freedom.

\section{Acknowledgments}
T.H. and M.B. acknowledge support from the German Research Foundation (DFG) within the
Project No. BI1694/1-1 and R.B. is thankful for funding from the GradUS program of the Saarland University.

T.H. and R.B. contributed equally to this work.

\end{document}